\documentclass[twoside,slac_one]{revtex4}
\usepackage{graphicx}
\usepackage{fancyhdr}
\usepackage{amsmath} % American Mathematics Society standards
\usepackage{bm}% bold math
\usepackage{amsxtra}
\usepackage{amssymb}
\usepackage{amsthm}
\usepackage{latexsym}
\usepackage{lscape}

\begin{document}

\title{Results and prospects for Charm Physics at LHCb}


\author{S. Borghi, on behalf of the LHCb collaboration}
\affiliation{Department of Physics and Astronomy, University of Glasgow, Glasgow, United Kingdom}

\begin{abstract}
Precision measurements in charm physics offer a window into a unique
sector of potential New Physics interactions. LHCb is well equipped to
take advantage of the enormous production cross-section of charm
mesons in $pp$ collisions at $\sqrt{s}=7$~TeV. The measurement of the
$D^0 -\bar{D}^0$ mixing parameters and the search for CP-violation in
the charm sector are key physics goals of the LHCb programme.  
The first CP violation measurements in the charm sector, 
 with 37 pb$^{-1}$ of data collected in 2010, are discussed.
The study of  $D^+ \rightarrow K^- K^+ \pi^+$ decays
shows no indication of CP violation.
The measurement of the proper time asymmetry in the time dependent
analysis of $D^0\rightarrow K^-K^+$ and $\bar{D}^0\rightarrow K^-K^+$ 
is evaluated to be $A_{\Gamma}=(-5.9 \pm 5.9_{stat} \pm 2.1_{syst})$.
The difference of CP asymmetry in the time integrated rates of
$D^0\rightarrow K^-K^+$ and $D^0\rightarrow \pi^- \pi^+$ decays
is measured to be $(-0.28 \pm 0.70_{stat} \pm 0.25_{syst}) \%$.
\end{abstract}

\maketitle

\thispagestyle{fancy}


%%%%%%%%%%%%%%%%%%%%%%%%%%%%%%%%%%%%%%%%%%%%%%%%%%%%%%%%%%%%%%%%%%%%%%%%
\section{Introduction}
LHCb \cite{lhcb}, an experiment at the Large Hadron Collider (LHC), is
dedicated to the study of  $b$ and $c$ flavour physics.  The
abundance of charm particles produced in LHC offers an unprecedented
opportunity for high precision measurements in the charm sector,
including measurements of  CP violation and $D^0 - \bar{D}^0$ mixing.
The high performance of LHCb detectors allows this potential to be
fully exploited.

The detector is a single-arm forward spectrometer covering
the geometrical region where heavy flavour particles, at LHC energy,
are mostly produced. A silicon micro-strip vertex detector
(VELO) provides, with high precision, the position of the primary vertex
and those of the decay  of long-lived particles. Other elements
of the LHCb tracking system include a silicon strip detector (TT) located in
front of a dipole magnet and three station detector downstream of the
magnet, composed of a silicon micro-strip detector (IT) in the inner part and
by straw drift chambers (OT) in the outer region.  Charged hadron
identification is made through two ring-imaging Cherenkov detectors (RICH).
The calorimeter system identifies high transverse energy hadron,
electron and photon candidates and provides information for the
trigger.  The particle identification system is completed by five muon
stations that provide fast information for the trigger and muon
tagging.

CP violation in $D$ decay processes has not yet been observed.  In the SM,
indirect CP violation in the charm sector is expected to be highly
suppressed, less than $\mathcal{O}(10^{-3})$, and universal between CP
eigenstates.  While, direct CP violation can be larger in SM dependent on
the final state:  CKM dynamics can produce direct CPV asymmetries in
single Cabibbo suppressed $D^{\pm}$ decays of the order of $10^{-3}$
or less \cite{theoryMS}.  Both asymmetries can be enhanced by New
Physics in principle up to  $\mathcal{O}(1\%)$ \cite{theoryNP}.

In 2010 LHCb recorded a total integrated luminosity of 37
pb$^{-1}$. This provides a charm sample large enough to be able
already to make several competitive measurements. The expected
integrated luminosity of more than 1 fb$^{-1}$ foreseen in 2011 will
offer the opportunity to improve the world knowledge of $D$ mixing and
CP violation.
The results of search for direct and indirect CP violation 
in the charm sector on data taken in 2010 by LHCb are presented here.
In particular, the search for direct CP violation in singly Cabibbo
suppressed (SCS) decay $D^+ \rightarrow K^- K^+ \pi^+$, 
the measurement 
of indirect CP violation in $D^0$ mixing in two body hadronic charm decays, and
the search for CP asymmetry
in the time integrated rates of $D$ mesons into 2 body SCS decays are illustrated.

%%%%%%%%%%%%%%%%%%%%%%%%%%%%%%%%%%%%%%%%%%%%%%%%%%%%%%%%%%%%%%%%%%%%%%
\section{Search for CP violation in $D^+ \rightarrow K^- K^+ \pi^+$ decays}

An independent analysis is performed to search for direct CP violation in
the singly Cabibbo suppressed decay $D^+ \rightarrow K^- K^+ \pi^+$.
The search consists of a direct comparison between the $D^+$ and the
$D^-$ Dalitz plots on a bin-by-bin basis.  The Dalitz plot is divided
into bins and  for each bin a local $CP$ asymmetry variable is
defined:
\begin{equation} 
S^i_{CP} = \frac{N^i(D^+)- \alpha N^i
  (D^-)}{\sqrt{N^i(D^+)+\alpha^2N^i(D^-)}}\: \mbox{,} \: \mbox{ with } \: 
\alpha=\frac{N_{tot}(D^+)}{N_{tot}(D^-)} \mbox{,}
\label{eq1}
\end{equation} 
where $N^i(D^+)$ and $N^i (D^-)$ are the numbers of $D^{\pm}$
decays in the $i$th bin and $\alpha$ is the ratio between the total
$D^+$ and $ D^-$ yields. The parameter $\alpha$ is a correction to
account for global asymmetries that are constant across the Dalitz
plot.

\begin{figure}[!t]
\centering
\begin{picture}(238.,155.2)
  \put(0,0){\includegraphics*[width=80mm]{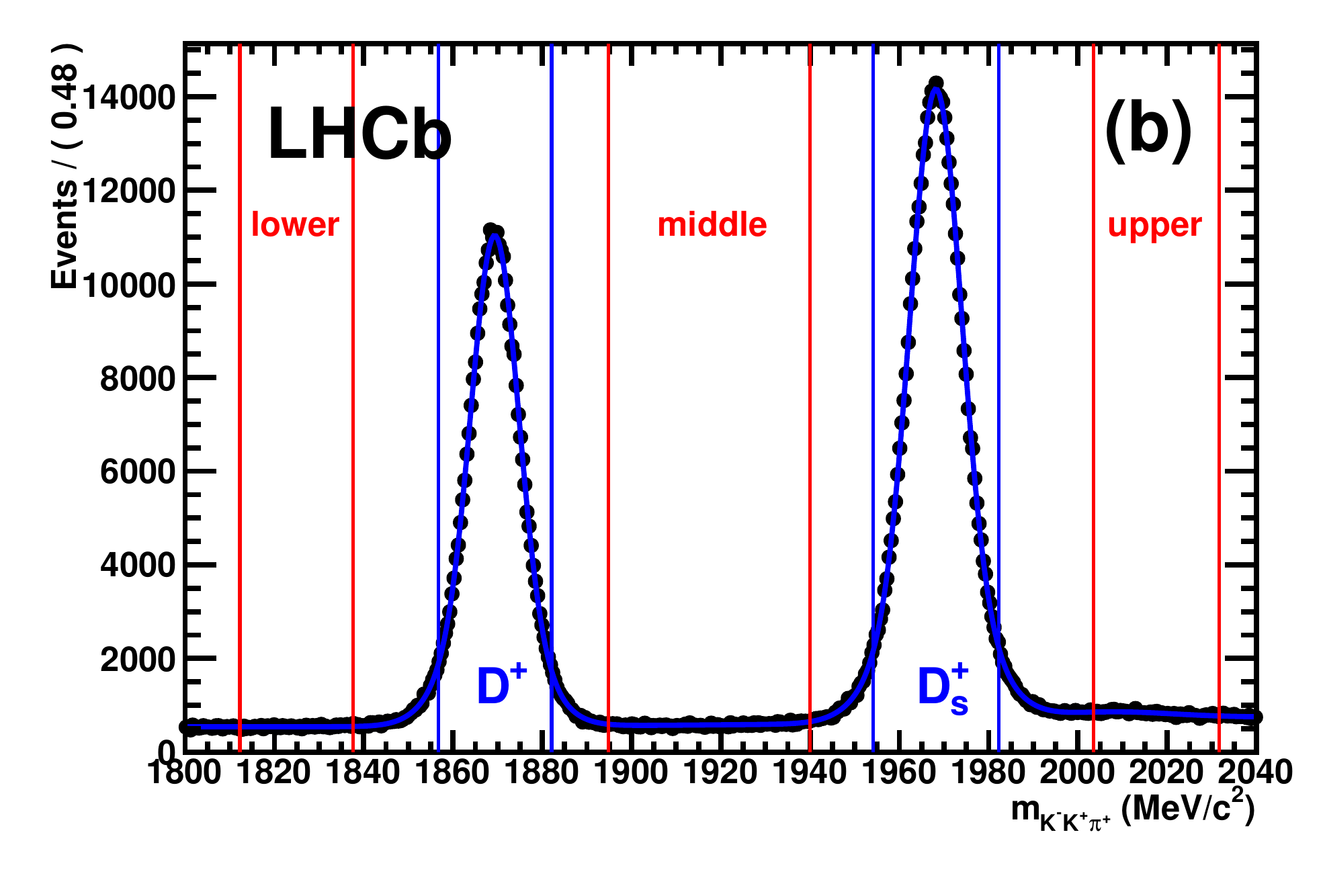}}
  \put(79.5,127){Preliminary}
\end{picture}
\caption{ Mass spectra of $K^-K^+ \pi^+$. The signal mass windows and sidebands (lower, middle and upper) are labelled.}
\label{mass3body}
\end{figure}

In the absence of local asymmetries, the $S^i_{CP}$ values are
distributed according to a Gaussian distribution with zero mean and
unit width. CPV signals are, therefore, deviations from this
behaviour.  The comparison between the $D^+$ and the $D^-$ Dalitz
plots is made by a $\chi^2$ test \cite{miranda}. The $\chi^2$ is defined as
$\chi^2 \equiv \sum(S^i_{CP})^2$ and the number of degrees of freedom ($ndof$) is the number of
bins minus one. Hence the probability value (p-value) measures
the confidence level that  the difference between $D^+$ and $D^-$ Dalitz
plots is driven only by statistical fluctuations.

Different binning schemes are considered to obtain the
highest sensitivity to various types of CPV.  The
bin scheme was optimized taking into account that we have no
sensitivity  if CP asymmetries change sign within a bin and we have a reduced
sensitivity if only a small part of a large bin has any CP violation
in it. 

The technique relies on careful accounting for local asymmetries that
could be induced by sources such as the different production
mechanisms for  $D^+$ and $D^-$, the difference in the K-nucleon
inelastic cross-section, differences in the reconstruction or trigger
efficiencies, left-right detector asymmetries, etc.
The existence of
these local asymmetries are investigated using the Cabibbo favoured
control channels, $D^+ \rightarrow K^- \pi^+ \pi^+$ and $D^+_s \rightarrow
K^- K^+ \pi^+$. No CP violation is expected in these channels.
The first
control mode, $D^+ \rightarrow K^- \pi^+ \pi^+$,  has an
order of magnitude greater branching ratio than the Cabibbo suppressed signal mode and
is more sensitive to detector effects since there is no cancellation
between $K^+$ and $K^-$.  The second control mode, $D^+_s \rightarrow
K^- K^+ \pi^+$, is similar to  the signal mode in terms of resonant
structure, statistics, kinematics, detector effects and backgrounds.
Similarly, the method is also applied in the sidebands 
(shown in Fig.~\ref{mass3body}) of the second control
channel to investigate possible asymmetries due to the 
contamination of the background.
 Another source of asymmetries could
come from a charge asymmetry from the parent $B$  in the $B\rightarrow
D(K^- K^+ \pi^+)X$ decays. The effect of secondary charm is
investigated by dividing the data set by the impact parameter\footnote{ The
IP is the minimum distance of approach with respect to the primary vertex.
The $\chi^2_{IP}$ is formed by using the hypothesis that the IP is equal to zero.}
 (IP) significance
($\chi^2_{IP}$) to have samples with different contamination from
secondary charm. 
All these tests are fully consistent with no asymmetry, thus
the method is determined to be very robust against systematic effects.

\begin{table}[!b]
\caption{The p-values for consistency with no CPV for the  $D^0 \rightarrow K^-K^+\pi^+$  decay mode
  for data with different magnet polarities}
\begin{center}
\begin{tabular}{|l|c|}
\hline \textbf{Magnet Polarity} & \textbf{p-value} \\ \hline Up &
6.0\%\\ \hline Down & 28.5\%  \\ \hline Combined & 12.7\% \\ \hline
\end{tabular}
\label{dalitztable}
\end{center}
\end{table}

The data sample used in this analysis corresponds to approximately 35
pb$^{-1}$ collected in 2010.
The signal sample consist of about 370k candidates.
The global asymmetries parameter ($\alpha$) is measured to be $0.984 \pm 0.003$. 
Fig.~\ref{mass3body} shows the invariant mass of  $K^-K^+\pi^+$ 
 for the analysed data sample. The
following two bin schemes are used on the signal data:  the first
scheme (Uniform) uses an uniform grid of equal size bins; the second
type (Adaptive) takes into account the non uniform event distribution
due to the $\phi \: \pi^+$ and $ \bar{K^*}(892)^0K^-$ modes. This second
scheme has bins of variable size, aiming for a uniform 
population in all the bins. For each bin the significance $S_{CP}^i$ of the difference
in $D^+$ and $D^-$ population is computed as defined in Eq. \ref{eq1}
and the $\chi^2/ndof$ is calculated to obtain the p-value.  The data
with opposite magnet polarities are combined to cancel left-right
asymmetries. 

\begin{figure}[!t]
\centering
\begin{picture}(238.,165.2)
  \put(0,0){\includegraphics*[width=80mm]{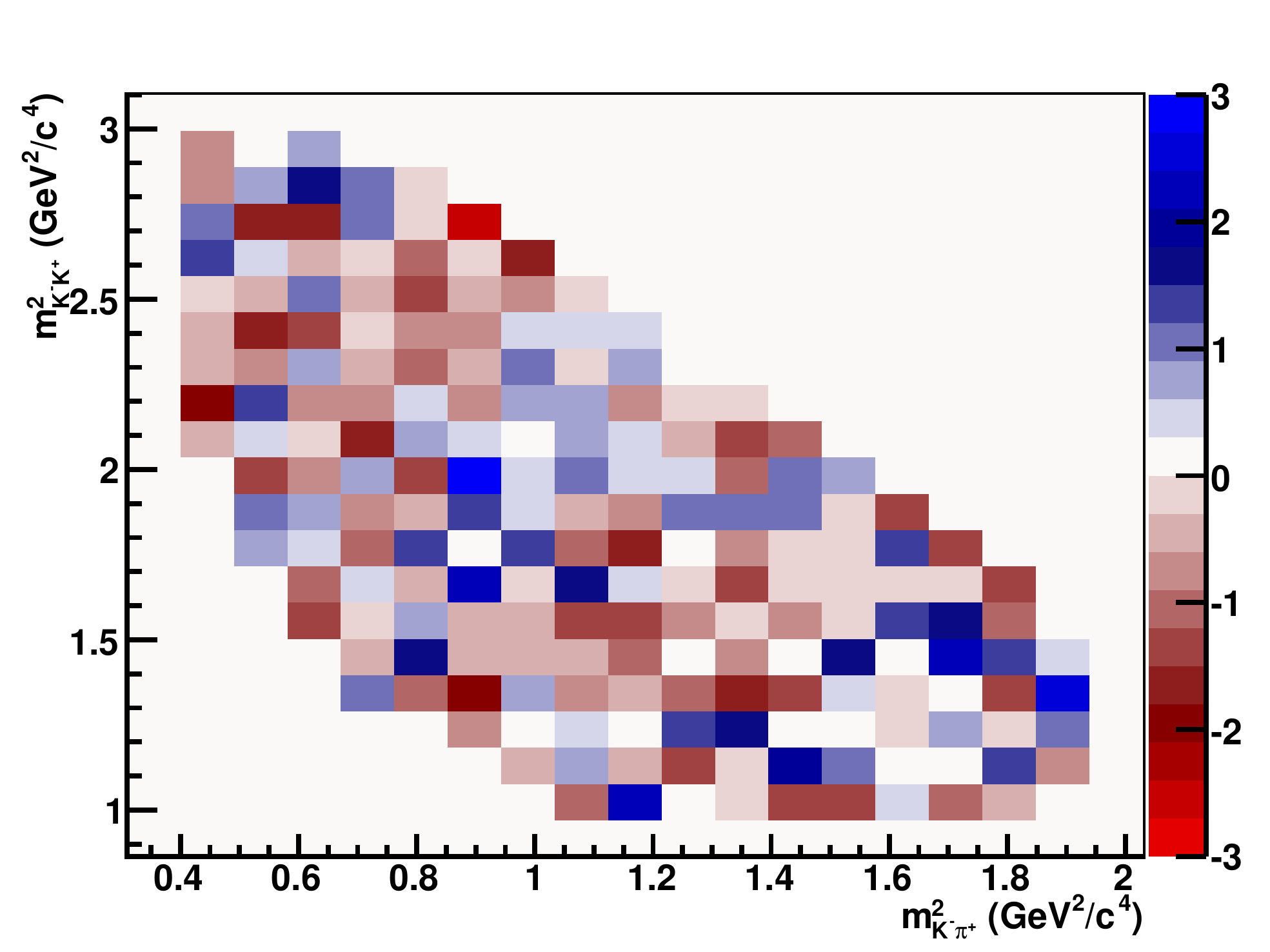}}
  \put(109.5,127){LHCb Preliminary}
\end{picture}
\caption{Distribution of $S_{CP}^i$ in the Dalitz plot for the Uniform scheme.}
\label{result3body}
\end{figure}

The obtained p-values, summarized in Table \ref{dalitztable},
 indicate no evidence for CPV \cite{3body}. This  result is also
supported by the result of the Gaussian fit of $S_{CP}^i$ that have
mean and width consistent with 0 and 1, respectively.
The results for the Uniform bin scheme are shown in 
Fig.~\ref{result3body}.  

%%%%%%%%%%%%%%%%%%%%%%%%%%%%%%%%%%%%%%%%%%%%%%%%%%%%%%%%%%%%%%%%%%%%%%%%
\section{Measurement of indirect CP violation in $D^0$ mixing}

A measurement of the indirect CP violation in $D^0$ mixing can be
performed in the study of two-body hadronic charm decays.  It can be
evaluated by the asymmetry of the proper-time ($\tau$) of flavour-tagged
decays:
\begin{equation} A_{\Gamma} \equiv \frac{\tau \left( \bar{D}^0\rightarrow K^-K^+
  \right)-\tau \left( D^0\rightarrow K^-K^+ \right)} {\tau \left(
  \bar{D}^0\rightarrow K^-K^+ \right)+\tau \left( D^0\rightarrow
  K^-K^+ \right)} = \frac{1}{2}\: A_M \: y \: \cos \phi - x \: \sin \phi \mbox{,}
\end{equation} 
where $x=\frac{\Delta m}{\Gamma}$ and  $y=\frac{\Delta
  \Gamma}{2\Gamma}$ are the mixing parameters and $\phi$ is the
CP violating weak phase.  $A_M$ is defined by the parameterization
$R_m^{\pm2} = \mid q/p \mid^{\pm 2} =1 \pm A_M$ with the assumption
that $R_m$ is close to unity and where $q$ and $p$ are parameters that
define the mass eigenstates in terms of the flavour eigenstates.

A measurement of $A_{\Gamma}$ differing significantly from zero would be
a measurement of indirect CP violation as it requires a non-zero value
for $A_M$ or $\phi$.

The signal yield and the background contribution are extracted from 
fits to the reconstructed invariant
mass alone. Due to the abundance of charm decays, the selection has been designed
to achieve maximal purity, with a background rate of the order of 
a few percent.
The main component of the background is due to the secondary charm,
i.e. $D$ mesons produced from $b$ hadron decays.  This kind of
background is not distinguishable by the invariant mass distribution.
The secondaries have larger impact
parameter with respect to the primary vertex than the prompts as a secondary $D$ no
longer has to point back to the primary vertex.  
Thus  this background can be reduced by a selection  based on the topology
but it can not be completely suppressed. Hence
a statistical separation is required. We use the variable
 $\mbox{ln} \left( \chi^2_{IP} \right)$, because it is an easier 
quantity to parameterise than $IP$ directly.
For the secondary charm, the $IP$ depends on the $B$ flight distance. 
The form of the consequent  $\mbox{ln} \left( \chi^2_{IP} \right)$ dependence 
on proper-time is extracted from the simulation, and the parameters of this 
dependence are evaluated in the fit procedure.

Since this analysis is sensitive to the proper time dependence of the
acceptance, particular attention is paid to requirements that could
bias this distribution.  A correction of these lifetime biasing effects
is  needed to properly extract $A_{\Gamma}$ via absolute lifetime
measurements. 
The heavy flavour selection implies some criteria which bias the
measured proper time distribution.  These biasing selections are
unavoidable and have to be applied already at trigger level to
suppress background from the large number of particles produced
promptly in the proton-proton collisions.  One example in the LHCb
selection of heavy flavour decays is the request of a minimum impact
parameter  of the daughters.  A correction of
these lifetime biases has to be computed to properly evaluate the lifetime to
determine $A_\Gamma$.
\begin{figure}[!t]
\centering
\includegraphics[width=80mm]{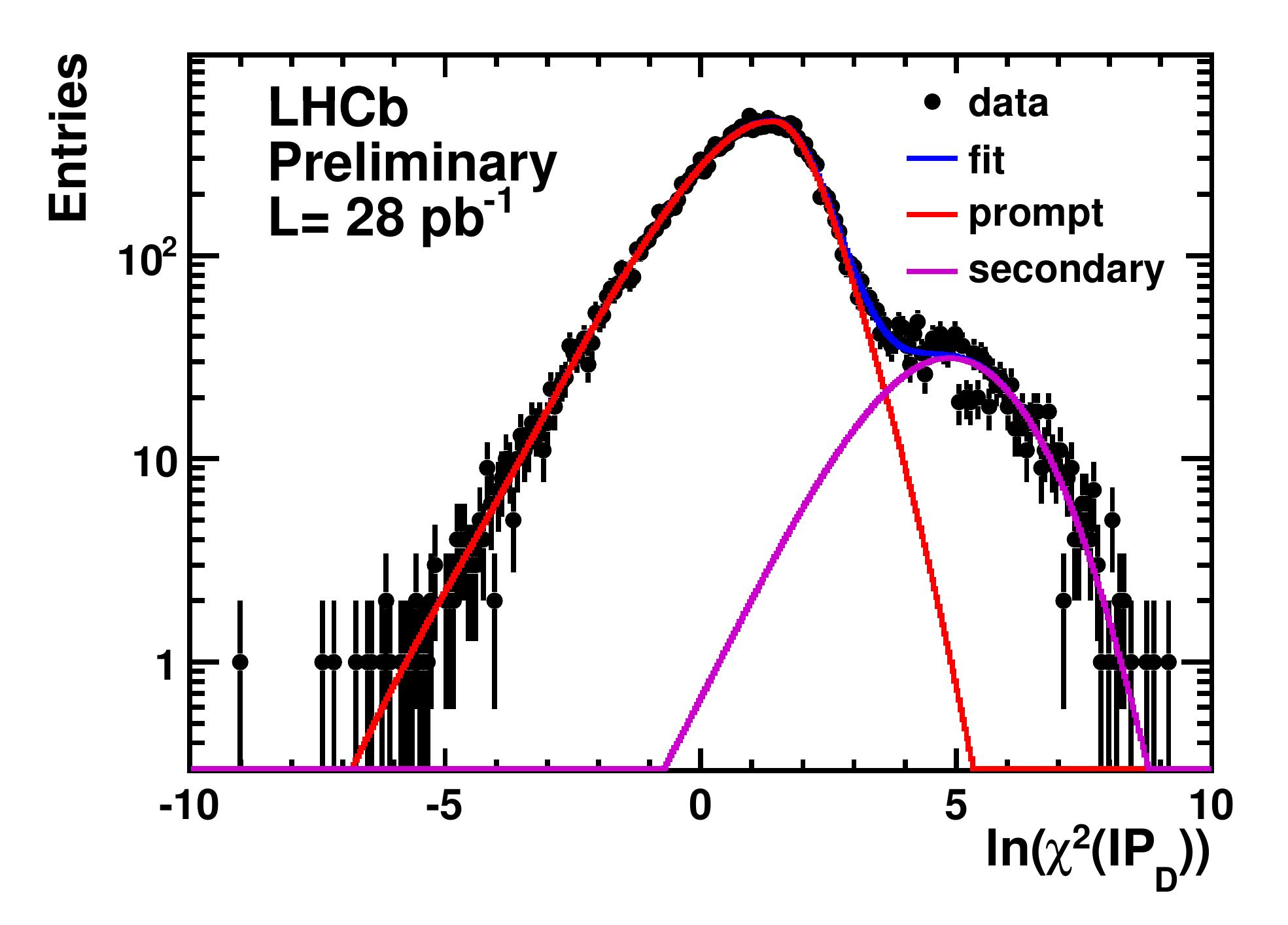}
\caption{$\mbox{ln} \left( \chi^2_{IP} \right)$ fit projection of $D^0\rightarrow K^-K^+$
  candidates. The data are shown as points, the total fit
  (blue), the prompt signal (red), and the secondary signal (pink).}
\label{ip}
\end{figure}

This analysis uses a data driven approach to evaluate the proper time
acceptance, that describes the selection efficiency as a function of the
$D^0$ proper time.  The method evaluates the proper-time acceptance on
a per event basis by a so-called `swimming' algorithm, which was
originally developed at CDF \cite{cdfswimming,cdfswimming2} and is now being
applied at LHCb \cite{swimming,swimming2,swimming3}.  In this method the acceptance
function is evaluated by moving the primary vertices (and thus varying
the proper time for the $D^0$ candidate). For each primary vertex
position  the software trigger decision and the offline selection are
re-evaluated. Consequently the proper-time acceptance function for
each event is determined as a sum of step functions, indicating when
the event would be selected or not selected.

The measurement of $A_{\Gamma}$ is performed via absolute lifetime
measurements obtained by a simultaneous fit of proper time and  $\mbox{ln}
\left( \chi^2_{IP} \right)$ including the acceptance function
evaluated by the swimming method.
Fig. \ref{ip} shows an example of the projection of the $\mbox{ln}(\chi^2_{IP})$ 
for $D^0\rightarrow K^-K^+$ decays. 

The measurement is based on a data sample equivalent to $28 \: \pm \:
3 \:$pb$^{-1}$ of data taken in 2010. The number of candidates selected
is about 15k for each flavour tag, $D^0$ and $\bar{D}^0$.
The flavour tagging of $D^0$ decays is done by reconstructing the
decay $D^{*+}\rightarrow D^0 \pi^+_s$, where the charge of the slow
pion ($\pi_s$) determines the flavour of the $D^0$ at production. 

The method was validated on a control measurement using decays to the
Cabibbo favoured decay $ D \rightarrow K \pi$. 
The combinatorial background contribution is $\sim 1\:\%$ for the
control channel and $\sim 3\: \% $ for $D\rightarrow KK$ decays.  
In the 2010 data sample only low statistics were available in the mass sidebands. 
Hence it was not possible to model the background shape, and the background contribution is
 neglected in the
time dependent fit and it is taken into account in the systematic
uncertainties.

In the control channel the result for the lifetimes, 
averaged between $D^0\rightarrow K^- \pi^+$ and  $\bar{D}^0\rightarrow K^+ \pi^-$,
is
$\tau(D^0) = 410.3 \pm 0.9 \: \mbox{fs} $,
where the uncertainty is statistical only.
This is in agreement with the current world average \cite{agammaworld}.
The lifetime asymmetry has been determined as
$A^{K\pi}_{\Gamma} = (-0.9 \pm 2.2_{stat} \pm 1.6_{syst})
\cdot  10^{-3} $, which is consistent with zero in accordance with the 
expectation.

\begin{figure}[!bt]
\centering
\includegraphics[width=80mm]{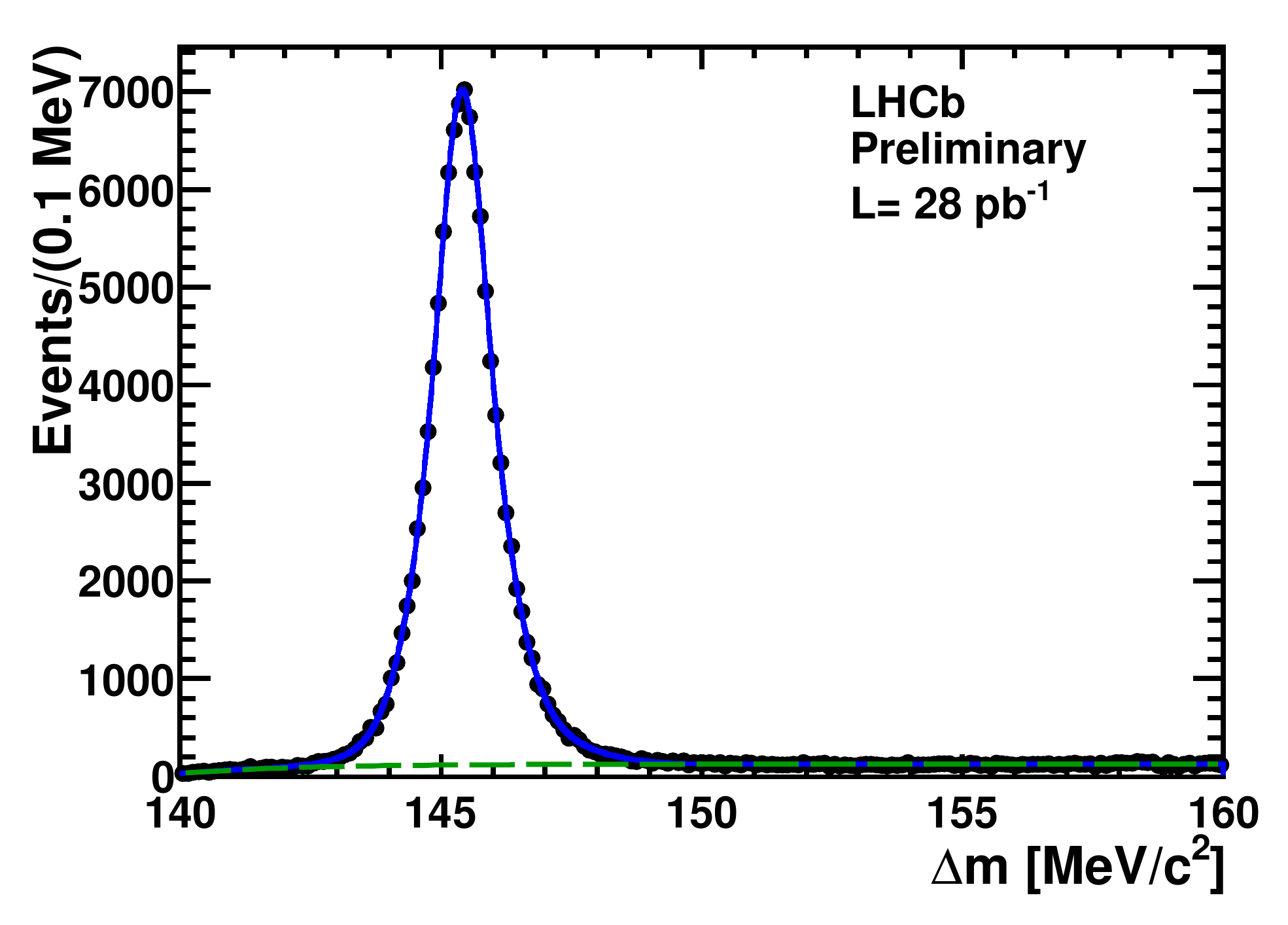}
\caption{$\Delta m$ mass difference between the reconstructed $D^*$ and $D$ candidates.
The data are shown as points, the total fit as a solid line and the
  random background as dashed line.} 
\label{deltam}
\end{figure}

The measured lifetime is an effective lifetime since the fitted
distribution includes also mistagged events, in which the $D^0$ is
associated with a random slow pion.  The mistag rates are 
assumed to be independent of the final state and are extracted from
the favoured $D\rightarrow K \pi$ decays which offer higher statistical precision.
This rate is evaluated by the fit of the difference between the mass of
 $D^*$ and $D^0$ ($\Delta m$) to be 1.8\%. This has been neglected 
 in the control channel as it is very
small, but it is applied in the evaluation of $A_{\Gamma}$.
The distribution of the $\Delta m$ between the $D^*$ and $D^0$ is shown in Fig. \ref{deltam}.
\begin{figure}[!t]
\centering
\includegraphics[width=80mm]{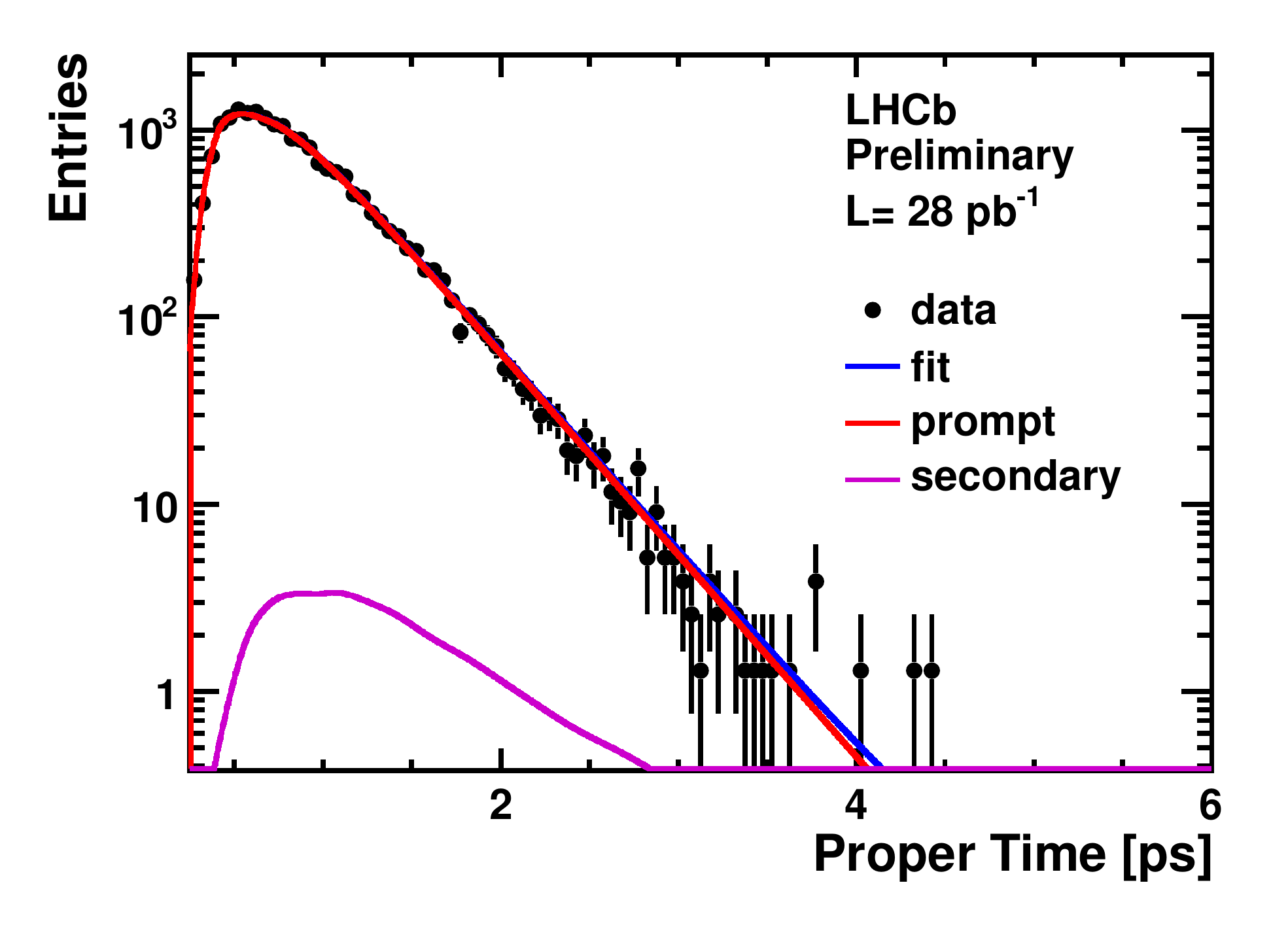}
\includegraphics[width=80mm]{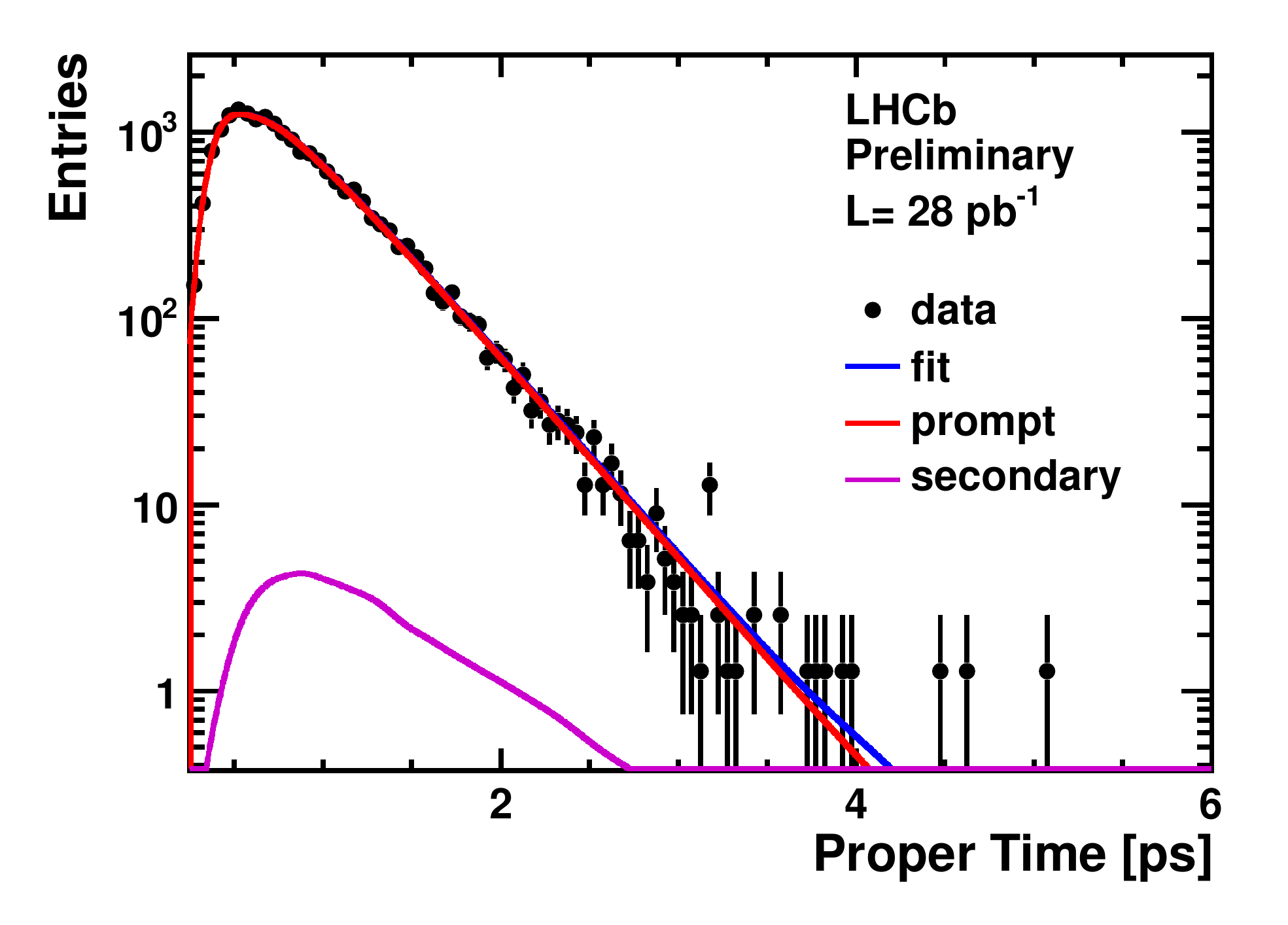}
\caption{Lifetime fit projection of $D^0\rightarrow K^-K^+$ candidates
on the left and $\bar{D}^0\rightarrow K^-K^+$ candidates
on the right  on a logarithmic scale. The data are shown as points, the total fit (blue), the
  prompt signal (red), and the secondary signal (pink).} 
\label{lifetime}
\end{figure}

The results of the lifetime fit of $D^0\rightarrow K^-K^+$ and 
$\bar{D}^0\rightarrow K^-K^+$
are shown in  Fig. \ref{lifetime}. 
The asymmetry is evaluated from these lifetimes to be  \cite{agamma}:
\begin{equation}
A_{\Gamma} = (-5.9 \pm 5.9_{stat} \pm 2.1_{syst}) \cdot 10^{-3} \mbox{.}
\end{equation} 
This result is consistent with zero and hence shows no evidence of CP violation
 and is in agreement with
the current world average \cite{agammaworld}.  
The main contributions to the systematic error are due to neglecting the
combinatorial background and to the separation of prompt
and secondary charm decays.
The systematic
uncertainty is expected to be significantly reduced by an improved
treatment of the background events, which will be possible for the data
taken in 2011.

%%%%%%%%%%%%%%%%%%%%%%%%%%%%%%%%%%%%%%%%%%%%%%%%%%%%%%%%%%%%%%%%%%%%%%%%%
\section{Search for CP asymmetry in the time integrated decay rates of $D$ mesons}
LHCb is searching for evidence of new sources of CP asymmetry in the
time-integrated decay rates of $D$ mesons.
The asymmetry is defined as
\begin{equation} A_{CP}^{RAW}(f)^* = \frac{N\left( D^{0} \rightarrow
  f\right)-N\left( \bar{D}^0 \rightarrow \bar{f}(\bar{f}\right)}
{N\left( D^{0} \rightarrow f\right)+N\left( \bar{D}^0 \rightarrow
  \bar{f}(\bar{f}\right)}
 = \frac{N\left( D^{*+} \rightarrow
  D^0(f)\pi^+\right)-N\left( D^{*-} \rightarrow
  \bar{D}^0(\bar{f})\pi^-\right)} {N\left( D^{*+} \rightarrow
  D^0(f)\pi^+\right)+N\left( D^{*-} \rightarrow
  \bar{D}^0(\bar{f})\pi^-\right)} \mbox{,}
\end{equation} 
where $N(X)$ refers to the number of reconstructed events of decay
$X$ after background subtraction.
The raw time integrated asymmetries of $D^0$ and $\bar{D}^0$ decays are considered
separately using the $D^*$ decay slow pion tagging method explained
above.

The raw asymmetries may be written as a sum of various components,
coming from both physics and detector effects:
\begin{equation} 
A_{CP}^{RAW}(f)^* = A_{CP}(f)+A_D(f)+A_P(D^*) \mbox{,}
\end{equation} 
where $A_{CP}(f)$ is the physics CP asymmetry, 
$A_D(f)$  the detection asymmetry of the $D^0$, $A_D(\pi_s)$
 the detection asymmetry of the soft pion and $A_P(D^{*})$ the production
asymmetry.

Taking the asymmetry difference of the two final state 
$\left( A_{RAW}(f)^* - A_{RAW}(f')^* \right)$ the production and soft
pion detection asymmetries will cancel. 
Moreover, for a two body decay of a spin-0 particle to a self-conjugate final
state, there is no $D^0$ detector efficiency asymmetry contribution,
i.e. $A_D(K^-K^+)=A_D(\pi^- \pi^+) = 0$.
Due to possible production and detection asymmetries, the measurement
of time-integrated CP asymmetry independently in $D^0\rightarrow KK $
and $D^0 \rightarrow \pi \pi$ is challenging.

We can however measure the difference in time-integrated CP asymmetry $\Delta
A_{CP}$ between  $D^0\rightarrow KK $ and $D^0\rightarrow \pi \pi $. 
\begin{eqnarray} 
A_{CP}^{RAW}(KK)^* = A_{CP}(KK)+A_D(\pi_s)+A_P(D^{*+}) \mbox{,}\\
A_{CP}^{RAW}(\pi\pi)^* = A_{CP}(\pi\pi)+A_D(\pi_s)+A_P(D^{*+}) \mbox{,}\\
\Delta A_{CP}^{RAW} = A_{RAW}(KK)^*-A_{CP}^{RAW}(\pi\pi)^* 
=A_{CP}(KK)+A_{CP}(\pi\pi) \mbox{.}
\label{eq8}
\end{eqnarray} 
No dependence remains on production or detection efficiencies, so
this observable is extremely robust against systematic biases.

In a proton-proton collider machine the production of heavy-flavour hadrons
need not be CP symmetric in a given region of phase
space. 
Possible variations of both selection efficiency and production and
detection asymmetry as a function of $p_T$ and $\eta$ could generate
second-order yield asymmetries that do not cancel out in our
formalism \ref{eq8}.
$A_{CP}^{RAW}$ extraction is performed in bins of $\eta$ and $p_T$ chosen
such that the statistics are approximately constant within each bin.
The binning is chosen to take into account the potential
variation of production or detection asymmetries in these variables
that differ for the two final states, such as those that may be induced in
the selection by e.g. particle identification requirements. 

\begin{figure}[!bt]
\centering 
\includegraphics[width=80mm]{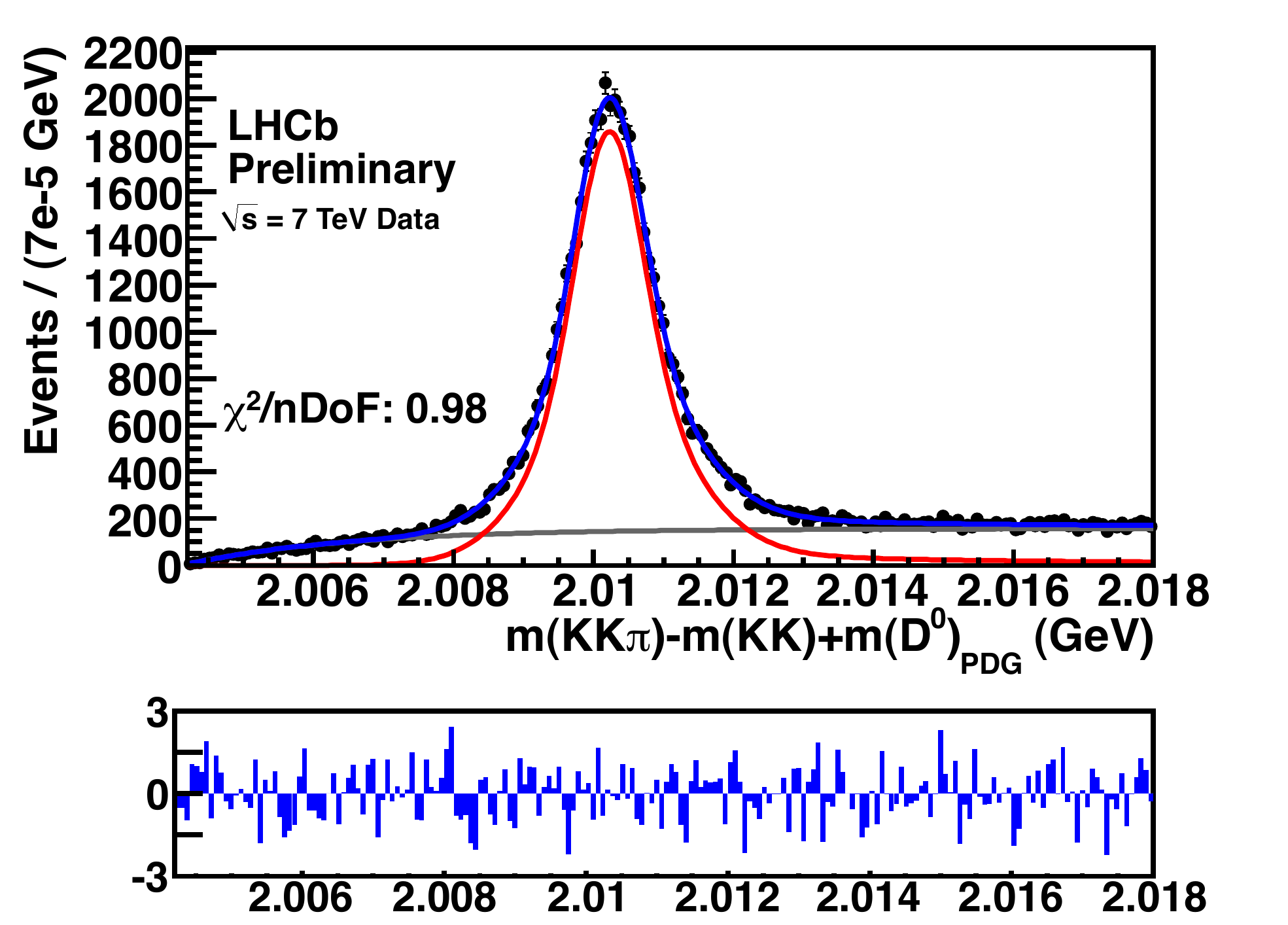}
\includegraphics[width=80mm]{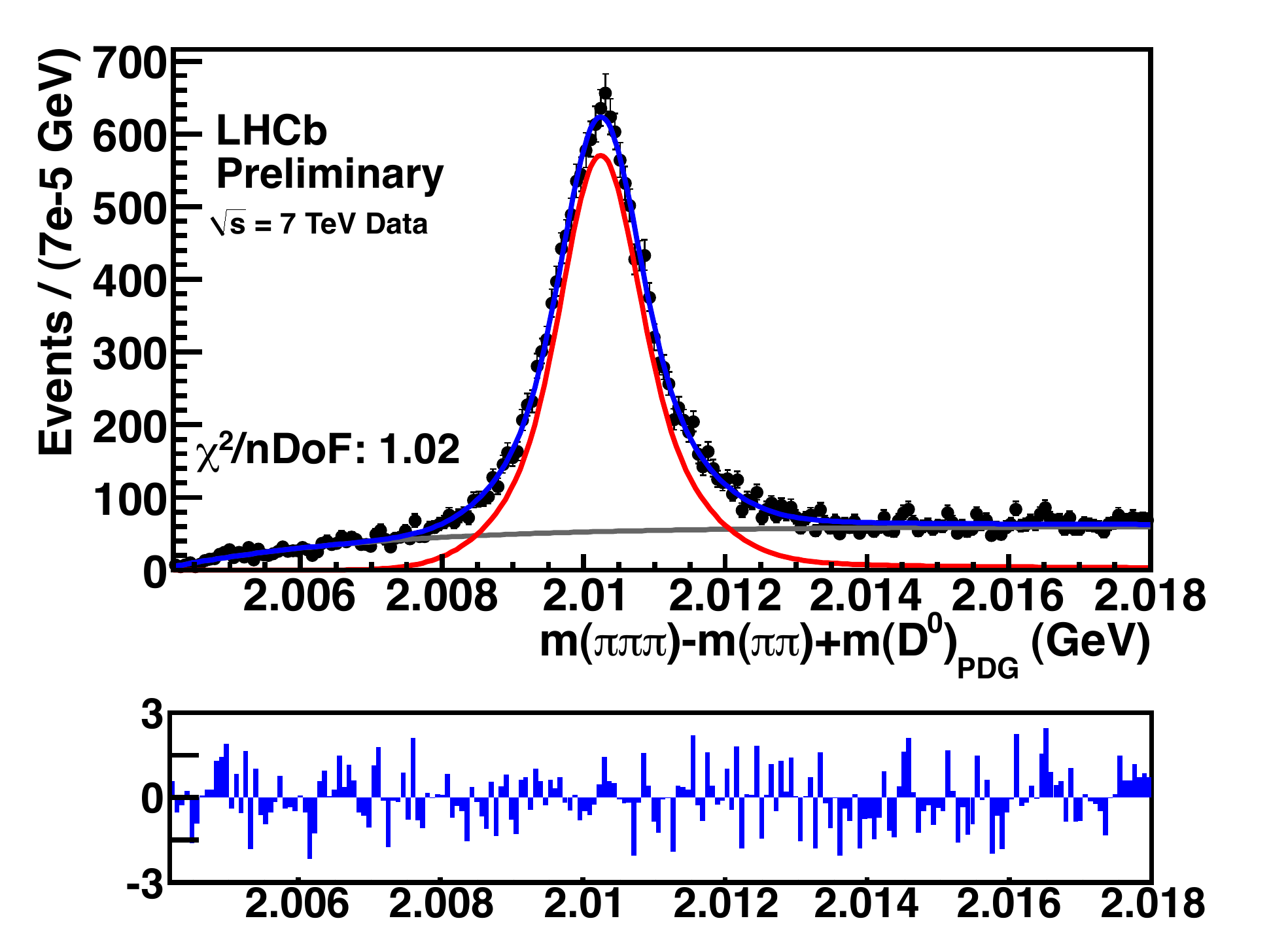}
\caption{Example fits to the mass difference spectra for tagged
  candidates for a subset of the data. The $D^0$ decay is
  reconstructed in the final states $K^-K^+$ (left) and $\pi^-\pi^+$
  (right). The normalized residuals for each bin are shown below.} 
\label{deltamACP}
\end{figure}

A binned maximum likelihood fit to the spectrum of the mass difference
between $D^*$ and $ D^0$ is used to evaluate the yields. Examples of the fit are 
shown in Fig.\ref{deltamACP}.
The data sample has an integrate luminosity of 37~pb$^{-1}$. The 
total signal yield is 116k tagged $D^0 \rightarrow K^-K^+$ and 36k
tagged $D^0\rightarrow \pi^- \pi^+$ .
The background of mis-reconstructed $D^*$ decays that peaks in the mass
difference is estimated from the mass sideband to be at the sub-percent
level. The effect enters the asymmetry calculation at second order,
$\mathcal{O}(10^{-4})$, and can be neglected.

Systematic uncertainties are
assigned by repeating the analysis with an alternative description of
the mass spectra lineshapes; with different fit windows for the $D^0$
mass; with all candidates, choosing one candidate randomly in events 
containing multiple candidates; and comparing with
the result obtained with no $(p_T, \eta)$ binning.  The full change in
result is taken as a systematic uncertainty and all uncertainties are
added in quadrature.  No source of limiting systematic bias has been
identified. These uncertainties are expected to be reduced by exploiting
the much larger statistics that will be available.

A value of $\Delta A_{CP}$ is determined in each measurement bin using
the result from $A_{RAW}(K^-K^+)^*$ and $A_{RAW}(\pi^-\pi^+)^*$.
These values are found to be consistent throughout the $(p_T, \eta)$
space, as well as for the two trigger periods and for both settings of
the magnet polarity. A weighted average is therefore performed to
yield the result  $\Delta A_{CP}= (-0.28 \pm 0.70 \pm 0.25) \%$  \cite{acp}.
This result is approaching the sensitivity of CPV measurements
performed by the B-factories in these decay modes\cite{bfactoryAcp, bfactoryAcp2},
but not yet at the level of CDF's recent measurement \cite{cdf}.

The time-integrated CP asymmetry $\Delta A_{CP}$ between the final states $D^0 \rightarrow
K^- K^+ $ and $D^0\rightarrow \pi^- \pi^+ $ has two contributions: a direct and an
indirect component, $a_{CP}^{dir}$ and $a_{CP}^{ind}$,
respectively. The indirect component may be assumed to be the same for
both final states as it originates in the common box diagram. However,
its time dependence has to be taken into account, leading to a
non-cancellation if the two final states are reconstructed with a
different mean proper time. While the direct component is different,
in general, for different final state.
Thus, the physics asymmetry of each final state may be written at first
order as \cite{acpdirind}
\begin{equation} \Delta A_{CP}\approx a_{CP}^{dir}(\pi^-\pi^+)+ \frac{\Delta
  <t>}{\tau} \: a_{CP}^{ind} \mbox{,}
\end{equation}
 where $\Delta <t>$ denotes the difference of the mean proper time
of the two final states and $\tau$ is the true $D^0$ lifetime.
Due to the difference in proper-time acceptance between the $K^-K^+$ and
$\pi^- \pi^+$ samples  $\Delta <t> / \tau = 0.10 \pm0.01$.
Although the measured value of $\Delta A_{cp}$ includes
a residual 10\% of the mode independent indirect CP asymmetry, 
 $\Delta A_{cp}$ is primarily sensitive
 to direct CPV. 
Thus the measurement of $\Delta A_{CP}$ and $A_{\Gamma}$ are complementary
in the search for CP violation. The current knowledge of these measurements
leads to an agreement with the no CP violation hypothesis 
with a C.L. of 20\% \cite{hfag}.

%%%%%%%%%%%%%%%%%%%%%%%%%%%%%%%%%%%%%%%%%%%%%%%%%%%%%%%%%%%%%%%%%%%%%%%%%%%%%%%%%%%%%%%%%%%%%%%%%%%%%%
\section{Conclusion}

The first measurements at LHCb for search for CP violation in the charm sector 
are competitive with the results of the B-factories, even though only a total integrated
luminosity of 37~pb$^{-1}$ collected in 2010 is used.
The search for direct CP violation  in $D^+ \rightarrow K^- K^+ \pi^+$ decays with a method
based on the study of Dalitz plots indicate no
evidence of CP violation.
The measurement of indirect CP violation is performed on singly Cabibbo suppressed  two-body
 charm decays through the study of the asymmetry of the proper-time. It is evaluated to be 
$A_{\Gamma} = (-5.9 \pm 5.9_{stat} \pm 2.1_{syst})$.
The time-integrated CP asymmetries of $D^0\rightarrow K^-K^+$ and $D^0\rightarrow \pi^- \pi^+$
decays is measured to be  $\Delta A_{CP}= (-0.28 \pm 0.70 \pm 0.25) \%$.
In addition to these measurements, many others are under way, e.g. in 2-body decays
 the measurement of the mixing 
parameters using doubly Cabibbo-suppressed $D^0\rightarrow K^+ \pi^-$ decays.
Significant improvements in the precision is expected with the large data set collected 
in 2011 with an expected integrated luminosity of about 1~fb$^{-1}$.

%%%%%%%%%%%%%%%%%%%%%%%%%%%%%%%%%%%%%%%%%%%%%%%%%%%%%%%%%%%%%%%%%%%%%%%%%%%%%%%%%%%%%%%%%%%%%%%%%%%%%%
\bigskip % extra skip inserted

\end{document}